# Multi-Agent System for Groundwater Depletion Using Game Theory

Ying Huang[1], Pavel Janovsky[2], Sanjoy Das[1], Stephen M. Welch[3], Scott DeLoach[2]
[1]Electrical & Computer Engineering Department
[2]Computer & Information Science Department
[3]Division of Agronomy
Kansas State University, Manhattan, KS 66505

## Abstract

Groundwater is one of the most vital of all common pool resources throughout the world. More than half of groundwater is used to grow crops. This research models groundwater depletion patterns within a multi-agent system framework. Irrigators are modeled as agents in the multi-agent system.

The irrigation strategies adopted by the agents are investigated using game theory, under several futuristic scenarios. The consequence of unregulated groundwater extraction in each case is analyzed. A set of five irrigators, growing three crops: corn, sorghum and wheat, have been considered in this study. To allow groundwater flow, these agents are assumed to be located in adjoining farm lands. Irrigators are modeled selfish agents that strategize their irrigation patterns in order to maximize their own utilities, i.e. the difference between the total revenue obtained from crop sales and the costs incurred, including groundwater extraction costs. Due to groundwater flow, irrigators have no incentive to conserve groundwater for later use. This leads to unsustainable depletion of the resource. Using the Nikaido-Isoda relaxation algorithm, their irrigation strategies under Nash equilibrium, when no irrigator can increase its utility by unilaterally changing its strategy, are obtained.

All parameters in this research are representative of Kansas. Recorded environmental and economic data of the region, along with the DSSAT software, have been used to obtain these futuristic projections. These scenarios include temperature increase, lowering of the water table, different precipitation levels, and different price increases for the crops. One of the emergent phenomena of the simulations is the adoption of crop rotation patterns by the irrigators to conserve groundwater. The irrigators grow corn, which is a more profitable yet water intensive crop in one year, and in the next, conserve water by growing sorghum instead. Another emergent outcome of this research is the viability of LEMAs. When the irrigators are subject to LEMA-level limits on groundwater use, there is a slight increase in the aggregate utility of the LEMA.

# 1. Introduction

## 1.1 Background

Groundwater is one of the most vital of all natural resources throughout the world that is currently being consumed at unsustainable levels throughout the globe [WB+10, SC+12]. Groundwater depletion threatens to be a major worldwide problem with severe social, economic and ecological consequences, including the drying up of wells, increased pumping costs, deteriorated water quality, and land subsidence. The long-term effects of water scarcity include famine, the outbreak of disease, and socio-economic and political conflict. Within the US, over half of the population relies on groundwater for drinking and other domestic use. Additionally, groundwater constitutes about 67% of all water used for irrigation [Bar14, MK+14].

The high rate of depletion of groundwater has been attributed to the observation that due to the tendency of the water to flow from one geographic location to another, there is little incentive for individual irrigators to conserve water. Hence their rational strategy is to overexploit the resource, leading to the situation commonly referred to as the *tragedy of the commons* [Har68, GS80]. Fortunately, more recent studies indicate that the coordinated use of groundwater can be very effective in addressing sustainability issues [Og93]. Recent studies on groundwater usage that take into account spatial hydrologic elements suggest that the coordinated use of groundwater can be very effective in limiting its depletion [BSZ10].

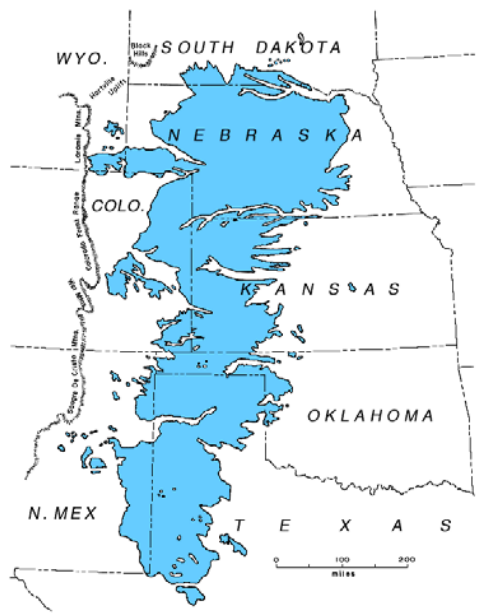

**Fig. 1.1.** The Ogallala aquifer (blue) in the Great Plains region of the US.
(Courtesy Kansas Geological Survey: www.kgs.ku.edu/Publications/Bulletins/162/gifs/fig006.gif).

The *Ogallala aquifer* is one of the most extensive bodies of water in the US, covering approximately 450,000 km$^2$ across South Dakota, Wyoming, Nebraska, Colorado, Kansas, Oklahoma, New Mexico and Texas [Lit09, Bar14, BKH14, MK+14, USG14, USG15], providing 30% of all groundwater extracted in the US and 82% of the drinking water demands of the region's 2.3 million strong population. An estimated $35 billion worth of crop yield each year relies directly on groundwater extracted from it [USG15].

Kansas' water users divert about 1.6–1.9 trillion gallons water for uses such as irrigation, power generation, public water supplies, industrial processes, stock-watering and other purposes, with the Ogallala aquifer being the primary source. About 85% of this water is used for irrigation [KWO93, MCG09, MG11, KDA13, KGS13]. Unfortunately, large-scale irrigation of the Ogallala aquifer, driven by technological innovations as well as increased demands of corn production for bio-fuels, has raised groundwater extraction to unsustainable levels [Wan11].

Future extrapolations of climate change, precipitation, and water needs present a bleak scenario for the Ogallala aquifer [HP11, Wan11, SB+13, BKH14]. A recent study projects that current extraction levels would result in groundwater depletion of 39% within the next 50 years [SB+13]. This could be further exacerbated based on recently published projections of the Central Plains region [CAS15]. Recharge supplies 15% of the current pumping level. At this rate, it would take an estimated 500–1,300 years to completely refill it.

## 1.2 Literature Survey

A significant amount of research on modeling approaches of natural resource sustainability, including groundwater is available in the existing literature [Fis12]. In order to address the conflicting goals of stakeholders in maximizing economic gains through groundwater extraction and eliciting sustainable consumption, several such studies have applied a multi-objective approach. One such study takes into account the objectives of water users, stakeholders, and a governmental agency (EPA) [KFB+10]. Six different water management policies are evaluated using different multi-criteria metrics and found to produce similar results [HH08]. Elsewhere, multi-objective evolutionary algorithms are applied to evolve optimal groundwater extraction policies [STL+09]. Another study, which also considers sustainable water use as an objective, models groundwater extraction in semi-arid conditions [GSS12]. Although multi-objective approaches offer the means to look into the tradeoff between economically profitable groundwater depletion and its sustainable depletion, they do not address how sustainability can be furthered without reducing immediate economic gains.

Several game theoretic models of groundwater use, with agents representing either individual users or user clusters, have been proposed recently [KBF+10, HKW11, Zuc11, LPE13, Hai14, Dol15, OY15]. Several studies are limited to two-agent static games [GS80, Mad10, CPME12, GPK+13, Hai14]. Other studies that aim to capture the behavior of a larger number of agents either apply principles of evolutionary game theory [NWV07, MKA12] or use other means to simplify the underlying model, such as structurally simple spatial agent configurations [SGB11]. Although analytical models are attractive from the standpoint of mathematical tractability, they rely on several simplifications and consequently, do not consider details such as spatial heterogeneity of the groundwater resource, annual trends in weather patterns, as well as various miscellaneous nonlinearities present in the real world. It has been argued that such assumptions are not adequate to capture more detailed aspects of groundwater depletion [ASS+12].

More computationally intensive approaches have been employed in large-scale models of common pool resource use [Bou05, CAT10, HBR10, NLRY14, AN14], including that of groundwater [BBM+07, LE10, OKH+10, CPME12, OKH12, MBY14]. These models are more effective in incorporating finer geophysical and economic details of the real world [An12]. Various aspects of groundwater use have been explored using multiple agent based computer simulations, such as water rights trading [BBM+07], water distribution policies [OKH12, OKH+10], and optimal economic allocation of the resource [MBY14, NLRY14].

Groundwater depletion in various regions of the world, including that taking place in the Maule river basin (Chile) [BBM+07], Jaguaribe basin (Brazil) [OKH12, OKH+10], Zamora aquifer (Mexico) [LPE13], and Andhra Pradesh (India) [ASS12] have been performed using various agent based models. However, to the best of the authors' knowledge, no such detailed models have so far been applied to the Ogallala aquifer.

A significant amount of research has been directed at exploring various methods to escape the tragedy of the commons. They have consistently revealed that cooperative groundwater extraction is instrumental in addressing sustainability [NWV07, Mad10, MD11, MKA12, CPME12, Hai14]. Unfortunately, methods that directly induce groundwater users to cooperate have largely remained unaddressed. The role of a central planner in coordinating groundwater use among agents has also been explored [SGB11, MBY+14]. Other studies reported in the recent literature on sustainable groundwater depletion include promoting environmental literacy among its users [LPE13], attaining sustainable consumptive practices through regulatory institutions [MD11, ASS+12], as well as penalizing overuse through payment schemes [Dol15] or taxation [RW13].

## 1.3 Research Contributions

The Ogallala aquifer provides a unique opportunity to study the behavior among irrigators under different scenarios, and the resulting impact on groundwater. Irrigators' usage patterns are driven through entirely economic considerations, with each irrigator trying to his or her own overall gain from that activity.

The research described here applies game theory to study the behavior of groundwater users (modeled as agents) under different scenarios, and how this collective behavior affects groundwater consumption. In contrast to earlier game-theoretic studies on groundwater resource use that adopt overly simplistic assumptions to derive game theoretic equilibrium, this research makes use of an iterative numerical technique to do so. Applying this computationally intensive process allows various real-world complexities to be incorporated into the proposed model. The simulations are carried out within a multi-agent systems (MAS) framework and over a period of several years, as well as under a variety of climatic as well as socio-economic time-varying conditions.

Implementing policies in order to elicit sustainable consumption is receiving much attention among policy planners, government and other administrative agencies. A recent example in groundwater was adopted by the state of Kansas in 2012, authorizing the formation of *Local Enhanced Management Areas* (LEMAs). A LEMA is a local group of users who enter legally enforceable limits on extraction for water rights within a specified geographic boundary. The research outlined in this thesis applies groundwater depletion constraints upon individual users operating under a LEMA. The results suggest that LEMAs can be very effective in eliciting sustainable groundwater use with little impact on the profitability.

# 2. Game Theory & Equilibrium

## 2.1 Non-Cooperative Game Theory

A competitive game consists of $N$ *agents* (or *players*) who participate in the game. Each agent can influence the outcome of the game by selecting a *strategy*. The strategy of each agent $i$ ($1 \leq i \leq N$) is denoted as $x_i$ and the set of all strategies available to agent $i$ as $X_i$ (so that $x_i \in X_i$).

The *joint strategy* is the vector of the strategies of all agents and is denoted as $\mathbf{x}$, so that $\mathbf{x} = (x_1, x_2, \ldots, x_N)$. Furthermore, as is commonly used in game theoretic notation, the collective strategies of all other agents excluding agent $i$ is denoted as $\mathbf{x}_{-i}$. In other words $\mathbf{x}_{-i} = (x_1, \ldots, x_{i-1}, x_{i+1} \ldots, x_N)$ and the joint strategy can be denoted as $\mathbf{x} = (x_i, \mathbf{x}_{-i})$.

The net gain of an agent from participating in the game is called its *utility*. The utility $U_i$ of any agent $i$ is a function that is determined not only by the strategy that that agent takes, but also by the strategies selected by all other agents in the game. Therefore $U_i: X_1 \times X_2 \times \ldots X_N \to \mathcal{R}$ ($i = 1, 2, \ldots, N$). It is convenient to express utilities explicitly as functions of the strategies, either as $U_i(x_i, \mathbf{x}_{-i})$ or as $U_i(\mathbf{x})$.

## 2.2 Nash Equilibrium

In a non-cooperative game, agents are assumed to be *selfish* as each agent tries to maximize its own utility from the game, while disregarding the utilities of all other agents. Under this situation, with the strategies of the other agents $\mathbf{x}_{-i}$ held constant, the agent $i$ will select a *best response* strategy $x_i$ as per the following expression,

$$x_i = \underset{y_i \in X_i}{\mathrm{argmax}}\, U_i(y_i, \mathbf{x}_{-i}). \tag{2.1}$$

Agents can change their strategies in any arbitrary order, in accordance with Eqn. (2.1). Nash equilibrium arises when no agent can improve its utility by unilaterally deviating from its previous its strategy. Thus a joint strategy $\mathbf{x}^* = (x_1^*, \ldots, x_N^*) = (x_i^*, \mathbf{x}_{-i}^*)$ is Nash equilibrium if for each agent $i = 1, 2, \ldots, N$ the following conditions hold,

$$U_i(y_i, \mathbf{x}_{-i}^*) \leq U_i(\mathbf{x}^*). \tag{2.2}$$

Looking at it another way, if $\mathbf{x}^*$ is at Nash equilibrium, then the best response of each agent $i$ is the agent's own Nash equilibrium strategy $x_i^*$ itself,

$$x_i^* = \underset{y_i}{\mathrm{argmax}}\, U_i(y_i, \mathbf{x}_{-i}^*). \tag{2.3}$$

## 2.3 Nikaido-Isoda Relaxation

Consider the two joint strategies $\mathbf{x} = (x_1, x_2, \ldots, x_N)$ and $\mathbf{y} = (y_1, y_2, \ldots, y_N)$. From $\mathbf{x}$ if any agent $i$ were to unilaterally change its strategy from $x_i$ to $y_i$, the new joint strategy would be $(x_1, \ldots, y_i, \ldots, x_N)$. This is denoted as $y_i | \mathbf{x} = (x_1, \ldots, y_i, \ldots, x_N)$. The increment in its utility due to the change in strategy would be $U_i(y_i | \mathbf{x}) - U_i(\mathbf{x})$. The *Nikaido-Isoda function* is defined as [CK24],

$$\psi(\mathbf{x}, \mathbf{y}) = \sum_{i=1}^{N} \bigl( U_i(y_i | \mathbf{x}) - U_i(\mathbf{x}) \bigr). \tag{2.4}$$

From any joint strategy $\mathbf{x}$ suppose each agent $i$ switches strategy to its best response, *i.e.* from $x_i$ to $y_i$, where, $y_i = \mathrm{argmax}_{y_i}\, U_i(x_1, \ldots, y_i, \ldots, x_N)$. Let us combine the separate best responses of each agent into a separate joint strategy $\mathbf{y} = (y_1, \ldots, y_i, \ldots, y_N)$. As each agent picks its best response to maximize its own utility, the Nikaido-Isoda function, $\psi(\mathbf{x}, \mathbf{y})$ is maximized. We define the *optimum response function* as the joint strategy that maximizes the Nikaido-Isoda function [CK24],
$$z(\mathbf{x}) = \underset{\mathbf{y}}{\mathrm{argmax}}\, \psi(\mathbf{x}, \mathbf{y}). \tag{2.5}$$

For any Nash equilibrium $\mathbf{x}^*$ (under certain conditions), the following must hold [CK04],
$$\mathbf{x}^* = z(\mathbf{x}^*). \tag{2.6}$$

The relaxation algorithm is shown below.

<u>Algorithm-2.1</u>
```
Initialize x⁽¹⁾
For l = 1 : ∞
    Obtain z(x⁽ˡ⁾) by 1-dim optimization for all agents
    x⁽ˡ⁺¹⁾ = (1 − η)x⁽ˡ⁾ + ηz(x⁽ˡ⁾)
End
```

Here $\eta \in (0,1)$ is a small constant. At the beginning of an iteration $l$ of the above relaxation algorithm, a new optimum response $Z(\mathbf{x}^{(l)})$ of the joint strategy $\mathbf{x}^{(l)}$ is determined. Next, the strategy $\mathbf{x}^{(l)}$ is incremented to bring it a little closer to $z(\mathbf{x}^{(l)})$. The updated strategy $\mathbf{x}^{(l+1)}$ is treated in an identical manner in the next iteration $l + 1$.

Under certain conditions, the above relaxation algorithm will converge to a Nash equilibrium $\mathbf{x}^*$,
$$\mathbf{x}^* = \lim_{l \to \infty} \mathbf{x}^{(l)}. \tag{2.7}$$
In practice, the relaxation process is carried out over a finite iterations $L_{\max}$, which while typically being small due to relatively rapid convergence of the relaxation algorithm, must be large enough so that only and the output $\mathbf{x}^{(L)}$ is significantly close to $\mathbf{x}^*$, i.e. $\|\mathbf{x}^{(L_{\max})} - \mathbf{x}^*\| \ll 1$.

# 3. Analytical Model of Groundwater Use

## 3.1 Groundwater

Groundwater, the common pool resource being modeled forms the backbone of this research. There are several factors affecting the groundwater head at any location. Groundwater is constantly supplied by local precipitation, which is the source of almost all freshwater in the hydrologic cycle [WHF98]. Precipitation includes rainfall as well as snow, although rainfall is the dominant source of local precipitation in the region considered in this study. Some of the locally available water is returned back into the hydrologic cycle by means of evaporation [LW60]. With $P$ denoting the precipitation per unit area, and $E$, the evaporation per unit area, our model combines both factors into a single quantity $R$, the net replenishment per unit area as shown below,

$$R(t) = P(t) - E(t). \tag{3.1}$$

Note that here and in the all other quantities involved in this investigation, the argument $t$ within parenthesis indicates the time step.

Groundwater is also subject to constant flow across locations. In the current framework, we have considered a total of $N$ agents, indexed as $i, j = 1,2, \dots, N$, that are involved in irrigation. Suppose $G_i$ and $G_j$ are the groundwater heads (in units of height) at the locations of agents $i$ and $j$, then the net groundwater flow from $i$ to $j$ is $a_{i,j}(G_j - G_i)$. The quantity $a_{i,j}$ is a physical constant corresponding to the rate of flow between the two locations. This quantity depends on the geology of the immediate region between the two locations, such as soil permeability, presence of rocks, etc. as well as elevations and area.

Under these circumstances, and in the absence of any human activity, the groundwater head at agent $i$'s location can be updated as,

$$G_i(t+1) = G_i(t) + R(t) - \sum_{j=0}^{N} a_{i,j}\left(G_j(t) - G_i(t)\right). \tag{3.2}$$

The summation term on the right in Eqn. (3.2) aggregates the groundwater flowing outwards from the location of agent $i$. Note that the summation is carried out over all $N$ agents but includes a quantity $G_0$ that represents the groundwater head in the surrounding region. Eqn. (3.2) is a straightforward implementation of the well-known Darcy's law [Dar56].

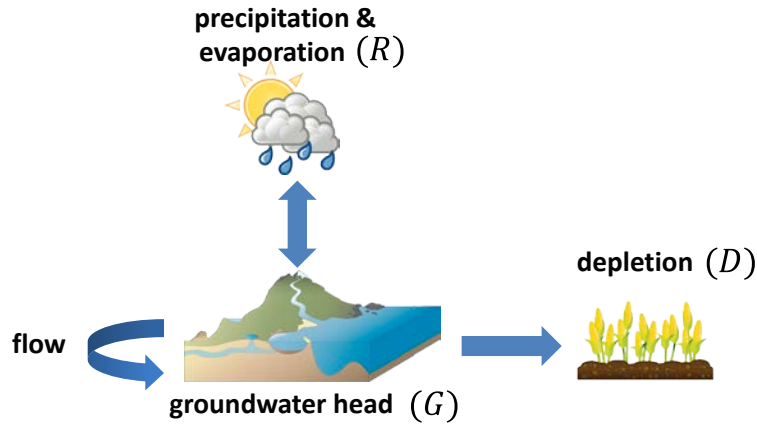

**Fig. 3.1.** Schematic showing all the factors involved in groundwater variation considered in Eqn. (3.3).

The addition of human activity to this model introduces another dimension affecting groundwater heads. This activity, carried out by the $N$ agents in the MAS framework in the form of irrigation, is a source of groundwater depletion and is also included in this model (see Fig. 3.1). With $D_i$ being the local groundwater depleted by an agent $i$ per unit area, the groundwater updating is now,

$$G_i(t+1) = G_i(t) + R(t) - D_i(t) - \sum_{j=0}^{N} a_{i,j}\left(G_j(t) - G_i(t)\right). \quad (3.3)$$

The surrounding groundwater is lowered linearly in terms of the time to model the steady drop in groundwater levels that is typically observed,

$$G_0(t+1) = G_0(t) - \gamma. \quad (3.4)$$

Here, $\gamma$ is the rate at which the average groundwater head is observed to drop in the geographic region under consideration.

## 3.2 Irrigation

In this model, each irrigator agent $i$ is assigned a fixed land area for irrigation. The agent can use this land to cultivate crops (indexed $k = 1, \dots, K$). Irrigation causes local groundwater to be depleted in two different ways. It is extracted directly by each irrigator agent through pumping in order to water the crop. Additionally, groundwater is also dissipated into the atmosphere through transpiration, the process of water continuously evaporating from the surface of leaf and released into the hydrologic cycle [LW60]. Both factors are dependent on the types of crop irrigated. Thus, the water depletion term in Eqn. 3.3 during a time period $t$ is given by,

$$D_i(t) = \sum_k (TR_k(t) + IR_k(t))x_{k,i}(t) \quad (3.5)$$

The quantities $TR_k$ and $IR_k$ pertain to the change in groundwater height from transpiration and due to groundwater being extracted for irrigation for crop $k$. The quantity $x_{k,i}$ in Eqn. (3.5) above is the fraction of the total land area that is used by agent $i$ to irrigate crop $k$. Note that $x_{k,i}$ must lie within the interval [0, 1], so that $0 \leq x_{k,i} \leq 1$. Other constraints placed on the variable $x_{k,i}$ are discussed later on in this thesis.

With $A_i$ being the total land area available for irrigation to agent $i$ the quantity of crop $k$ harvested at the end of the crop's farming season is given by,

$$Q_{k,i}(t) = A_i y_k(t) x_{k,i}(t). \quad (3.6)$$

The factor $y_k$ in the above product is the yield of crop $k$ per unit area. It depends on the environmental conditions during the entire irrigation period of that crop, from planting until the end of harvesting.

The gross revenue $U_i^R$ in monetary units that is obtained by the irrigator agent $i$ from crop sales is the sum of the revenue gained from the sale of each crop and given by the following expression,

$$U_i^R(t) = \sum_k p_k(t) Q_{k,i}(t). \quad (3.7)$$

In Eqn. (3.7), the quantity $p_k$ is the market price per bushel of the crop $k$.

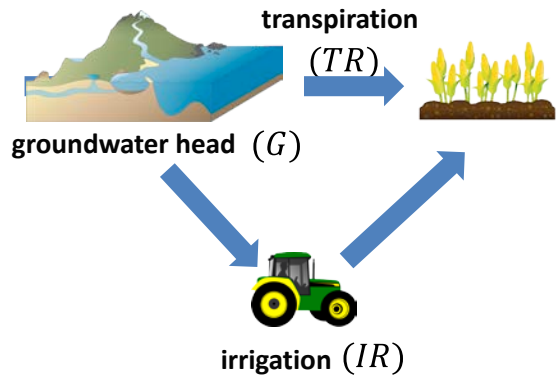

**Fig. 3.2.** Schematic showing the pathways leading to further groundwater depletion due to irrigation.

## 3.3 Investment

The irrigation of crops requires a certain amount of monetary investment arising due to groundwater extraction, as well as production costs. It is assumed that groundwater is extracted by means of gas operated pumps. The Nebraska Pumping Plant Performance Criteria provides a simple formula to calculate the corresponding pumping cost per unit area [RA99, DP12]. This expression is as follows,

$$EP_i(t) = \frac{\theta}{p} g(t)(G - G_i(t) + 2.31\psi). \quad (3.8)$$

In the above equation, the constant $\theta$ is the amount of natural gas required to lift a unit volume of water per unit height difference, $p$ is the pumping efficiency, $g(t)$ is the price per unit of natural gas and $\psi$ is the gauge pressure of the irrigation system which is multiplied by an appropriate conversion factor of 2.31.

The total cost of extracting groundwater by agent $i$ is given by,

$$U_i^E(t) = A_i EP_i(t) \sum_k IR_k(t) x_{k,i}(t). \quad (3.9)$$

The cost associated with the production of per unit area of crop $k$ is $c_k$. Since the crop is irrigated in a fraction of the total area, the area under irrigation of crop $k$ is $A_i x_{k,i}$ and the associated production cost is $c_k(t) A_i x_{k,i}(t)$. The total production cost associated with all crops is the sum of the production costs of all crops and is given by,

$$U_i^P(t) = \sum_k c_k(t) A_i x_{k,i}(t). \quad (3.10)$$

## 3.4 Game Theoretic Model

In the game theoretic model, a total of $K$ crops are considered. Each agent $i$ is involved in irrigating a fraction $x_{k,i}(t)$ of each crop $k = 1, \ldots, K$ during a time interval $t$ in years. The simulation is carried out in discrete time steps until $T_{\max}$, so that $t = 1, 2, \ldots, T_{\max}$. The strategy $\mathbf{x}_i$ of each agent $i$ is a $K \times T_{\max}$ vector consists of the fraction of its available area used to irrigate each crop, and during each time interval.

Two assumptions have been made for the sake of simplicity: (*i*) *fully informed agents*, and (*ii*) *utility time-invariance*. The first assumption implies that the agents can simultaneously access all environmental and economic future forecasts during all $T_{\max}$ time steps while optimizing their strategies. This assumption is made here in order to study the model without introducing irrelevant complexities arising from data uncertainties. When there is a significant amount of uncertainty (as in climate change), simulations have been carried out over multiple scenarios.

Under the second assumption of utility time invariance, the agents place equal weights to their net monetary gain for each time step $t = 1, \ldots, T_{\max}$. The rationale behind this simplification arises from several preliminary simulations that fortuitously showed that there the resultant changes in the agent's strategies when future utilities were progressively discounted by 2% - 5% each year were not large enough to change the conclusions drawn from this entire study. Discounts higher than that are rendered meaningless when $T_{\max}$ is kept at large values since discounted utilities well into the future are lower than floating point representation bounds.

This net gain of an agent $i$ at time $t$ can be computed readily as the difference between the revenue $U_i^R(t)$ obtained from Eqn. (3.7) and the costs $U_i^E(t)$ and $U_i^P(t)$ which are determined using Eqns. (3.9) and (3.10). The overall utility that the agent maximizes is the aggregate of all such gains and is given by,

$$U_i(\mathbf{x}_i, \mathbf{x}_{-i}|\mathbf{\Omega}) = \sum_{t=1}^{T_{\max}} U_i^R(t) - \left(U_i^E(t) + U_i^P(t)\right). \quad (3.11)$$

All input data to the model are included in $\mathbf{\Omega}$ in Eqn. (3.11).

The Nash equilibrium strategies analogous to Eqn. (2.3) and the corresponding utilities are,

$$\mathbf{x}_i^* = \underset{\mathbf{x}_i}{\operatorname{argmax}}\, U_i(\mathbf{x}_i, \mathbf{x}_{-i}^* | \mathbf{\Omega}), \tag{3.12}$$

$$U_i^* = \underset{\mathbf{x}_i}{\max}\, U_i(\mathbf{x}_i, \mathbf{x}_{-i}^* | \mathbf{\Omega}) = U_i(\mathbf{x}_i^*, \mathbf{x}_{-i}^* | \mathbf{\Omega}). \tag{3.13}$$

Note that each opponent joint strategy $\mathbf{x}_{-i}$ contains $(N - 1) \times K \times T_{\max}$ entries as it accounts for all remaining $N - 1$ agents other than $i$, and the joint strategy of all agents $\mathbf{x}$ includes a total of $N \times K \times T_{\max}$ real numbered entries. Fortunately, mathematical software packages (such as MATLAB used in this research) support multidimensional arrays.

With $\mathbf{x} \triangleq (\mathbf{x}_i, \mathbf{x}_{-i})$ and $\mathbf{y} \triangleq (\mathbf{y}_i, \mathbf{y}_{-i})$ being any pair of strategies, the Nikaido-Isoda function is given by,

$$\psi(\mathbf{x}, \mathbf{y}) = \sum_i \big( U_i(\mathbf{y}_i, \mathbf{x}_{-i} | \mathbf{\Omega}) - U_i(\mathbf{x}_i, \mathbf{x}_{-i} | \mathbf{\Omega}) \big). \tag{3.14}$$

The Nikaido-Isoda relaxation procedure specific to this model is shown below.

<u>Algorithm-3.1</u>
```
Initialize x
while (‖z(x) − x‖∞ ≥ ϵ) do
      z(x) = argmax ψ(x, y)
              y,c(y)=T
      x = (1 − η)x + ηz(x)
end
```

It should be noted that the Nash equilibrium is not unique in our model; initializing the relaxation algorithm with different values through different seeds the MATLAB random number generator produced different strategies under equilibrium. Similar observations have been reported elsewhere [Mad10]. Fortunately, there were no discernible differences in the agents' utilities across each simulation. All results provided throughout the remainder of this thesis have been obtained by simulating the Multi-Agent System (see Chapter 4) with the same random number seed value.

# 4. Multi-Agent System

## 4.1 Geographical Setting

The rate of groundwater decline has been estimated to be approximately one foot every year [USG13]. This rate of decline has been assumed in all simulation results reported in this thesis.

The study has been carried out with a set of five hypothetical irrigators located in Garden city in southwest Kansas. Garden city is located at $37°\ 58'\ 31''\ N\ 100°\ 51'\ 51''\ W$ as shown in Fig. 4.1 and its average elevation is 2,838 feet. The immediate neighborhood around this city is predominantly agricultural, with the adjoining Ogallala aquifer being the source of water needed for irrigation. Garden city is projected to face severe water shortage in future [Lit09].

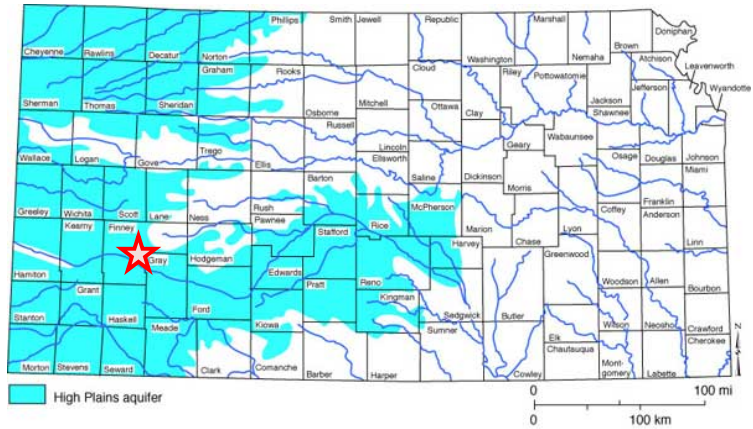

**Fig. 4.1.** Map of Kansas showing the extent of the Ogallala aquifer region (blue) and Garden city (red star).

## 4.2 Irrigation

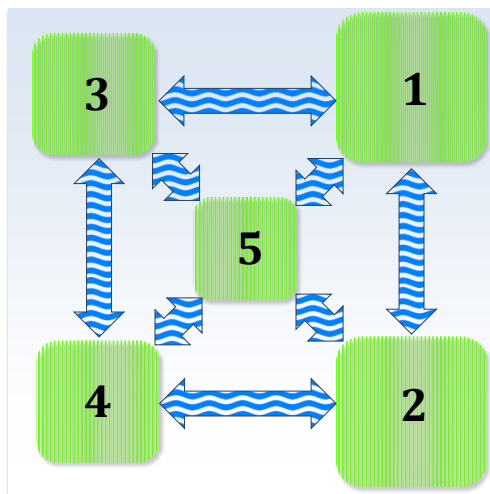

**Fig. 4.2.** Groundwater flow between the agents' locations. Box sizes reflect actual areas available to agents 1–5.

**Table 4.1.** Area and initial groundwater heads of each agent.

| Agent index $i$ | Area in acres $A_i$ | Initial ground-water head in meters $G_i(0)$ |
|---|---|---|
| 1 | 1200 | 125.0 |
| 2 | 1200 | 113.0 |
| 3 | 1000 | 125.0 |
| 4 | 1000 | 113.0 |
| 5 | 900 | 118.0 |
| 0 (Aquifer) | - | 118.8 |

Three crops have been considered in this research ($K = 3$), which are corn ($k = 1$), sorghum ($k = 2$), and wheat ($k = 3$). The study involves five irrigator agents ($N = 5$). Table 4.1 shows the area associated with each irrigator agent and their initial groundwater heads. Fig. 4.2 shows the permeability coefficients of the groundwater flow between the agents. The annual rate of drop of the aquifer's groundwater head (see Eqn. (3.4)) is kept at $\gamma = 304.8$ mm/year. This value is the average drop rate within the period.

As corn and sorghum are summer crops and wheat is a winter crop, agent strategies are constrained in the following manner,
$$0 \leq x_{1,i}, x_{2,i}, x_{3,i} \leq 1,$$
$$x_{1,i} + x_{2,i} = 1. \quad (4.1)$$

## 4.3 Physical Data Preparation

Decision Support System for Agrotechnology Transfer (DSSAT), a widely used software package with an underlying crop simulation model, has been used to generate crop related data for this research [D15]. The raw inputs to DSSAT consists of (*i*) weather data (**W**), (*ii*) crop data (**C**$_k$), and (*iii*) soil data (**S**). The package's outputs, for each crop $k$ each year, that are used here are the following: (*i*) the transpiration, $TR_k$, (*ii*) the water extracted for irrigation, $IR_k$, (*iii*) the evapotranspiration $ET_k$, (*iv*) the precipitation, $P_k$, and (*v*) the crop yield per unit area, $y_k$. Note that $ET_k$ and $P_k$ pertain to evapotranspiration and precipitation only during the relevant season that the crop is grown. The inputs and outputs of DSSAT are shown in Fig. 4.3.

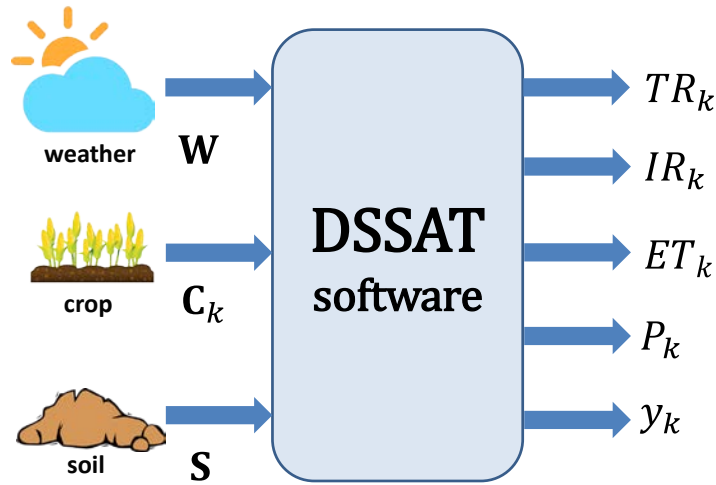

**Fig. 4.3.** Inputs and outputs of DSSAT; the outputs serve as input data during the MAS simulation.

Historical weather data between years 1964 – 2014 used as the raw inputs **W** to DSSAT are available by National Center for Environmental Information [NCD15]. This data consists of daily records of the precipitation, solar radiation, and the minimum and maximum temperatures. Aggregated over the entire year the precipitation data is used to obtain $P$ in Eqn. (3.1).

Data pertaining to Garden city, for each crop $k$ used in this study, which is available online at Kansas State University [CPT15] is used as the other input $\mathbf{C}_k$ to DSSAT. Finally, the soil data **S** was obtained from the US Department of Agriculture (USDA) [USDA15].

The evapotranspiration $ET_k$ obtained from DSSAT includes the transpiration $TR_k$ as well as the evaporation taking place during the growing season of each crop $k$. Hence the difference $ET_k - TR_k$ is the total evaporation during this season and independent of $k$. In this study, the average difference is taken as the evaporation that takes place within that season. In order to estimate the total water evaporation per unit area for the entire year $t$, it has been assumed for simplicity that the ratio of the precipitation during the growing season and that over the entire year is equal to the same ratio of the evaporation. In other words,

$$\frac{1}{P(t)}\sum_k P_k = \frac{1}{E(t)}\sum_k (ET_k - TR_k). \tag{4.2}$$

Eqn. (4.2) is used in this study to estimate the evaporation $E$ in Eqn. (3.1).

The outputs $TR_k$ and $IR_k$ from DSSAT summed across an entire year are used in Eqn. (3.5) and Eqn. (3.9). The output $y_k$ is used in Eqn. (3.6).

Unfortunately the historical input data to DSSAT, contains a great deal of variation across time that do not represent any long-term trends in the time series. These variations manifested themselves as 'noise' in the raw output data produced by DSSDAT. Initial experiments revealed that this noise level was large enough to subsume the differences across the different scenarios in some of our simulations. The approach taken was to express the outputs of DSSAT as a quadratic approximation $\widehat{\mathbf{\Psi}}_{\text{DSSAT}}()$ of its inputs that can be expressed succinctly in the following manner,

$$\begin{pmatrix} TR_k(t) \\ IR_k(t) \\ ET_k(t) \\ P_k(t) \\ y_k(t) \end{pmatrix} = \widehat{\mathbf{\Psi}}_{\text{DSSAT}}\big(\mathbf{W}(t = 1, \ldots, T_{\max}), \mathbf{C}_k(t = 1, \ldots, T_{\max})\big). \tag{4.3}$$

Note that the set of arguments of the approximating function $\widehat{\mathbf{\Psi}}_{\text{DSSAT}}()$ does not include the soil data **S** as one of its arguments. This is because in all our simulations this is a fixed quantity that does not vary with time and is implicitly incorporated within the coefficients of the quadratic function.

## 4.3 Socio-Economic Parameters

The raw market price $p_k$ of each crop $k$ was obtained from the National Agricultural Statistics Service [NAS15]. The raw prices were subject to exponential saturations to obtain the actual values used to compute the agent utilities in Eqn. (3.7). Accordingly, the market price $p_k$ of each crop $k$ is given by,

$$p_k(t) = p_k^\infty(t) + \big(p_k^0(t) - p_k^\infty(t)\big)e^{\frac{-\sum_i Q_{k,i}(t)}{\tilde{Q}_k}}. \tag{4.4}$$

In Eqn. (4.4), the saturated market price $p_k$ is bounded within the interval $(p_k^\infty \; p_k^0)$ and $\tilde{Q}_k$ is an appropriate scale constant. The summation term appearing in the exponent term's numerator is used to determine the total crop market supply. This treatment of the market prices is useful to imitate the limited market of sorghum.

The production cost $c_k$ for each crop was obtained from USDA [AERS15]. The cost of production $c_k$ of each crop $k$ is also subject to exponential saturation within the intervals $(c_k^\infty \; c_k^0)$.

$$c_{k,i}(t) = c_k^\infty(t) + \big(c_k^0(t) - c_k^\infty(t)\big)e^{\frac{-A_i x_{k,i}(t)}{\tilde{a}_k}}. \tag{4.5}$$

It can be seen that the quantity of the crop is not summed across all agents in Eqn. (4.5).

Owing to extraneous factors not considered in this research, all raw prices and costs contained significant variations. In order to capture only the long term variations, the raw data had to be replaced with trends,
$$\left(p_k^\infty(t), p_k^0(t)\right) = \left(p_k^\infty(0), p_k^0(0)\right)e^{\frac{t}{\tau_k}}. \tag{4.6}$$
Here the quantity $\tau_k$ is a time constant that minimizes the sum squared deviation from the observed data. In a similar manner, with $\theta_k$ being the analogous time constant, the costs were obtained as,
$$\left(c(t), c_k^0(t)\right) = \left(c_k^\infty(0), c_k^0(0)\right)e^{\frac{t}{\theta_k}}. \tag{4.7}$$
The source of raw gas price data $g$ was obtained from the U.S Energy Information Administration website [NGP15]. Upon exponential fitting, the gas price used in this study is,
$$g(t) = g(0)e^{\frac{t}{\zeta}}. \tag{4.8}$$

## 4.4 Simulation

The purpose of the MAS simulation is to determine the utilities $U_i(\mathbf{x}|\mathbf{\Omega})$ of each irrigator $i$ (Eqn. (3.11)). It serves as a procedure invoked by Algorithm 3.1. As each utility represents the aggregate over time, the MAS simulation, depicted below (Algorithm 4.1) is carried out sequentially from $t = 1$ until $T_{\max}$.

<u>Algorithm-4.1</u>
```
Input x
For each agent i  U_i = 0
t = 1
While t ≤ T_max
     Update G_i(t + 1) of each agent i
     Obtain DSSAT data TR_k(t), IR_k(t), ET_k(t), y_k(t)
     Compute U_i^R(t), U_i^E(t), U_i^P(t) of each agent i
     For each agent i  U_i = U_i + U_i^R(t) + U_i^E(t) + U_i^P(t)
     t = t + 1
End
Output U_i of each agent i
```

# 5. Results: Baseline Model

## 5.1 Setup

The first study of the proposed MAS is to explore how projected changes in future weather may produce shifts in local irrigation practices, and how this shift affects groundwater level. The next study of the proposed MAS is to explore how changes in social and economic parameters in the future will affect groundwater use. In order to perform comparative analyses, a baseline simulation is performed where all these parameters remain at constant values obtained from historical data. The prices of corn, wheat and sorghum are maintained at a constant level equal to the average prices during the ten-year period 2001 – 2010. Although the gas price shows an increasing trend during this period, it is also fixed at the average value. The simulation is carried out over a period of 20 years. All the inputs to the MAS are obtained using Eqn. (4.3) using the historical weather data **W** during the 20 year period 1995 – 2014.

## 5.2 Results

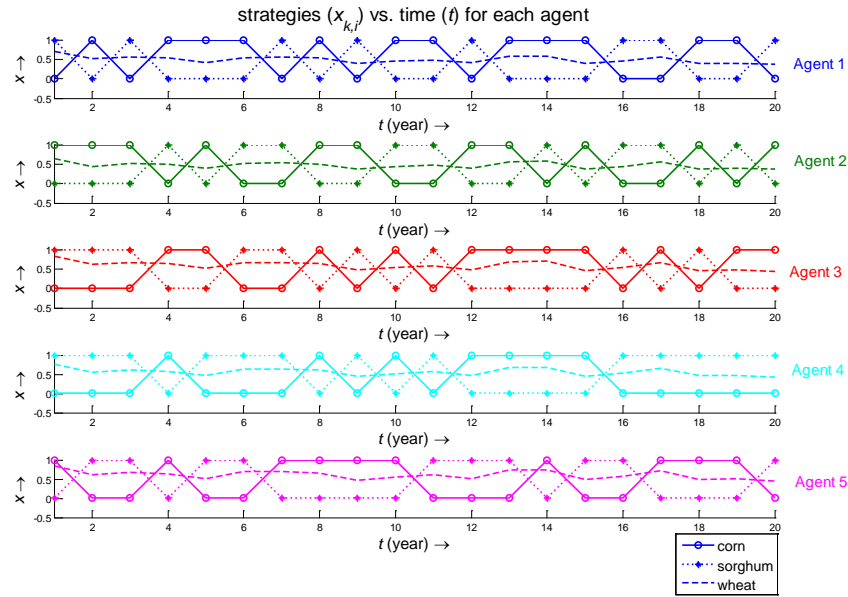

**Fig. 5.1.** Strategies vs. year for each agent (baseline).

Fig. 5.1 shows the strategies of the five irrigator agents evolving during the 20 year period under the baseline scenario. All agents used the entire land area for cultivation during summer, opting to grow either corn or sorghum. Although corn has a larger market than sorghum resulting in higher revenues, it also requires more water for irrigation. The tradeoff between the higher revenues ($U_i^R$) and lower extraction costs ($U_i^E$) is clearly seen in this figure. The agents switch regularly between the two crops, opting for higher revenues in some years and conserving water by growing sorghum during other years. It can be observed that the most common strategy is to switch between the two summer crops in alternate years, although choosing to grow the same crop for two (or even three) years is also seen. It can be seen that the initial groundwater head ($G_i(0)$) available with each agent influences its irrigation strategy. Although agents 1 and 2 have equal irrigation areas, as agent 1 has a higher it grows corn for 12 of the entire 20 yearly periods, while agent 2 grows corn for only 11 of those years. In a similar manner, although agents 3 and 4 have identical areas, they grow corn for 11 years and 7 years. Agent 5 which has an intermediate value of initial groundwater chooses to grow corn for 10 of the 20 years.

Figs. 5.2 and 5.3 show how the utilities and the utilities per unit area vary with time during this twenty-year period. In Fig. 5.2 it can be seen that, as expected, the utilities of agents 1 and 2, which have the largest areas under irrigation, are the highest, while that of agent 5 with the smallest area, is the lowest. When the utilities per unit area are compared, a similar pattern is observed, although the differences are not as discernible. Agents 1 and 2 have the highest utilities per unit area while agent 5 has the lowest. We hypothesize that this phenomenon reflects the economics of scale – irrigators with larger areas under cultivation are able to garner higher benefits from each unit of land.

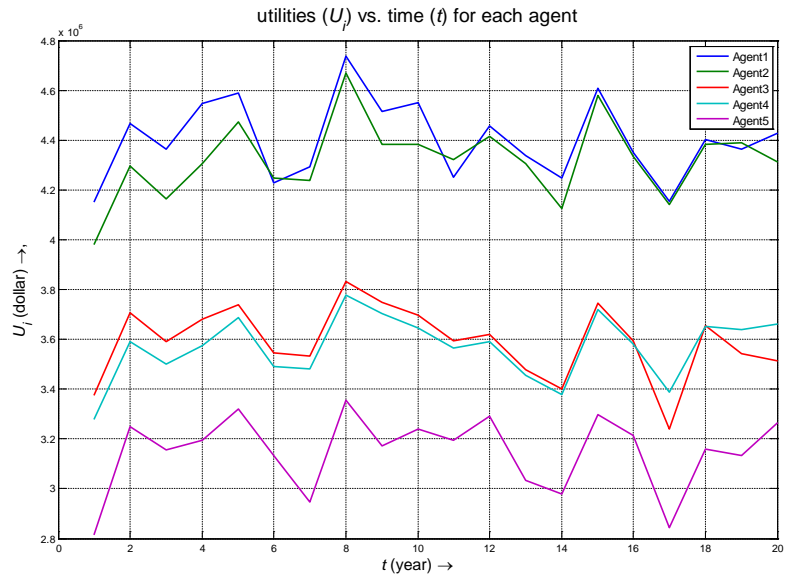

**Fig. 5.2.** Utilities vs. year for each agent (baseline).

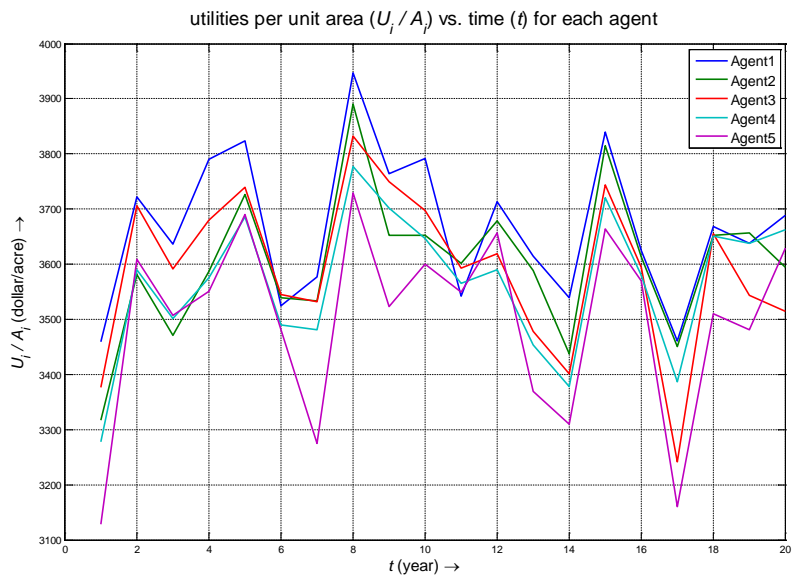

**Fig. 5.3.** Utilities per unit area vs. year for each agent (baseline).

The groundwater head at each agent's location as well as that of the surrounding aquifer is shown as a function of time in Fig. 5.4. It can be observed that constant irrigation throughout the entire simulation period for maximum utility clearly results in a larger decline in the groundwater in comparison to the groundwater head of the aquifer. It can also be seen that agents 1 and 3, which have higher initial groundwater levels show the steepest declines in their groundwater heads over time. In comparison, agents 2 and 4 show the least decline in comparison to the other agents. This observation can be explained by examining the amount of corn – the most water intensive, yet profitable crop, grown by each agent. Agents 1 and 3, have the highest groundwater heads at the beginning of the simulation, grow corn for a total of 23 years (see Fig. 5.1), averaging 11.5 years per agent. Agent 5, which begins with an intermediate water head, grows corn a total of 10 time periods, while agents 2 and 4 grow the crop for an average of only 9 years per agent. This pattern is consistent with earlier studies that report that higher groundwater availability leads the irrigators to exploit the resource rather than conserving it.

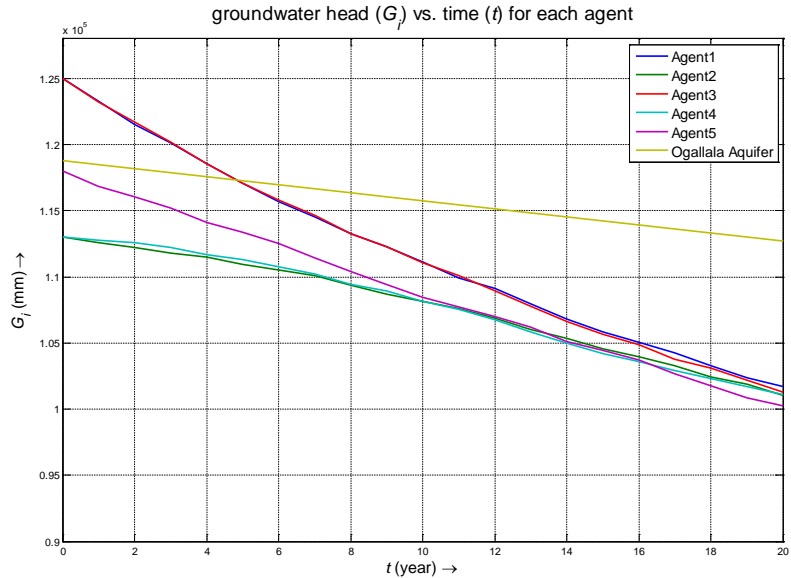

**Fig. 5.4.** Groundwater head vs. year for each agent (baseline).

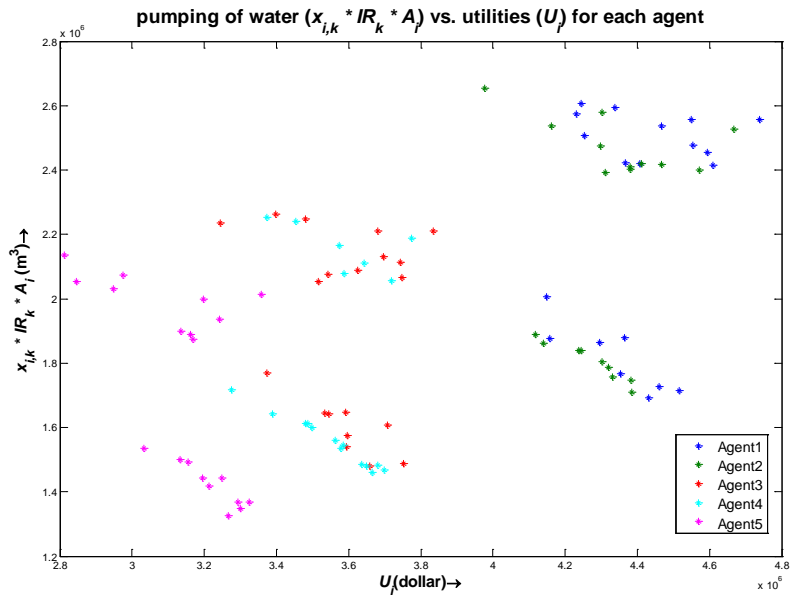

**Fig. 5.5.** Water pumped vs. utility for each agent (baseline).

In Fig. 5.5, the total volume of water pumped each year by each agent is plotted as a function of the agent's utility of that year. Upon visual inspection, it is clear that the points center around two distinct clusters per agent, depending on the summer crop the agent picks for the year.

# 6. Results: Future Scenarios

## 6.1 Climate Change Scenarios

There is near unanimity that the Great Plains region is expected to experience severe climate change during the mid 21st century, with large changes in rainfall patterns. One study predicts precipitation shifts towards wetter summers in the northern areas of the Great Plains by 10% whereas between 10% – 30% drier weather in the southern Great Plains [PC13]. Accordingly two possible scenarios are reported here that represent possible weather patterns during the period 2032 – 2051. The first scenario models a 20% increase in summer precipitation, while the other scenario considers a 10% uniform decrease in precipitation throughout the year. The historical precipitation data in **W** is changed accordingly to obtain the MAS inputs using Eqn. (4.3). Additionally as increased rainfall is associated with cloudier conditions, the summer solar radiance component of **W** is lowered by 5% for the first scenario, and uniformly increased by 2% in the other.

Fig. 6.1 shows the irrigation strategies of the agents under higher summer precipitation. Due to the increased availability of water, agents 1 and 2 grow corn for 18 of the 20 year simulation period. A similar upward trend in number of years used to grow corn is seen in the other agents. In Fig. 6.2, which shows the changed irrigation strategies with lower precipitation, the opposite trend is seen. Here, all five agents opt to cultivate corn for a smaller number of years.

The outcomes of the two future precipitation scenarios on the local groundwater levels are provided in Fig. 6.3 and Fig. 5.4. Although the rate at which groundwater drops in lower in Fig. 6.3, this change is not as significant as one would expect from a 20% increase in summer rainfall.

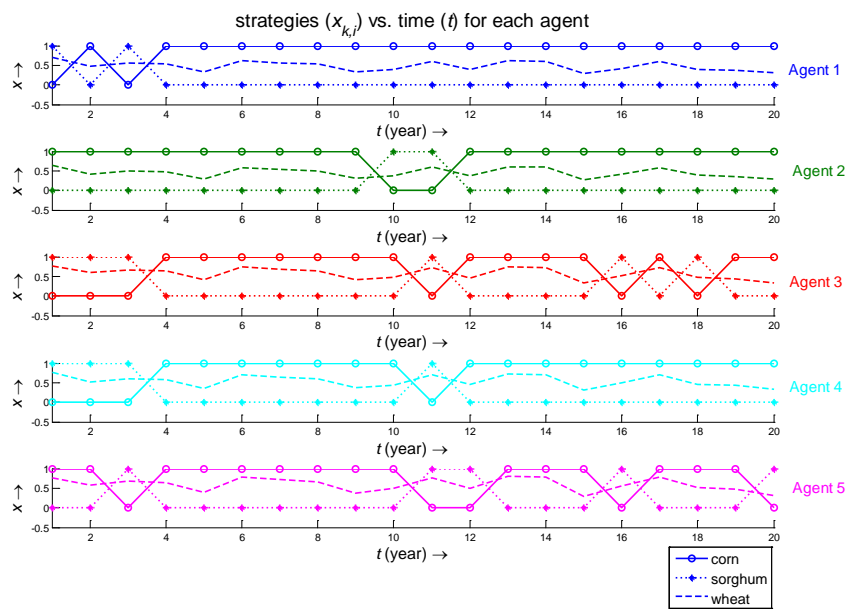

**Fig. 6.1.** Strategies vs. year for each agent (increased rainfall).

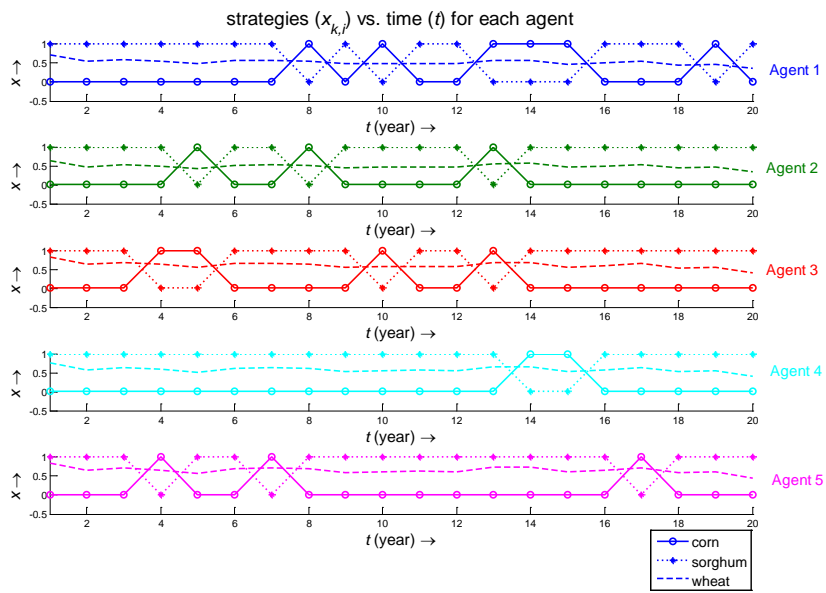

**Fig. 6.2.** Strategies vs. year for each agent (decreased rainfall).

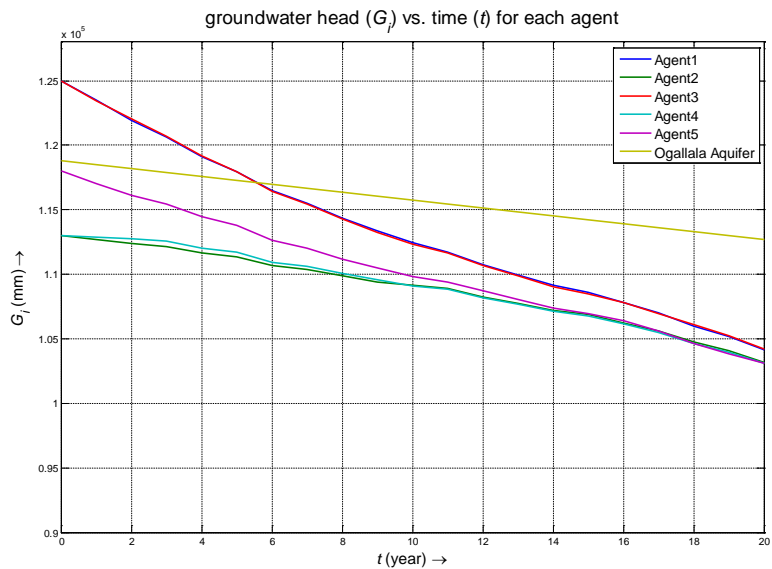

**Fig. 6.3.** Groundwater head vs. year for each agent and aquifer (increased rainfall).

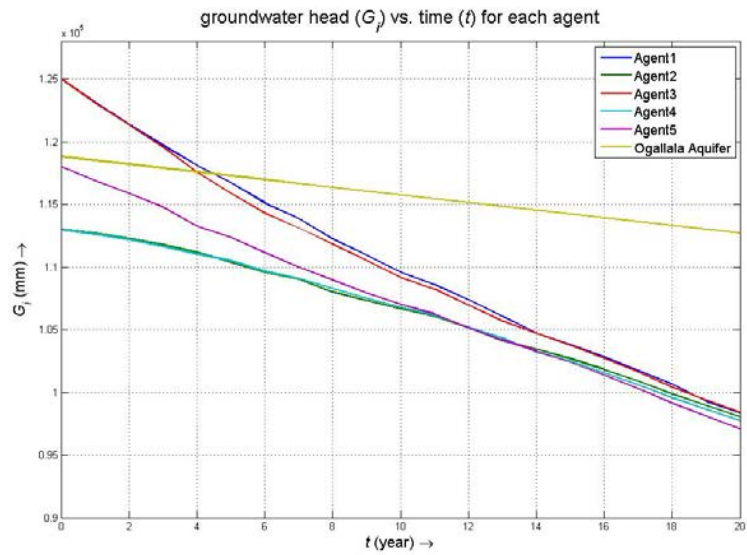

**Fig. 6.4.** Groundwater head vs. year for each agent and aquifer (decreased rainfall).

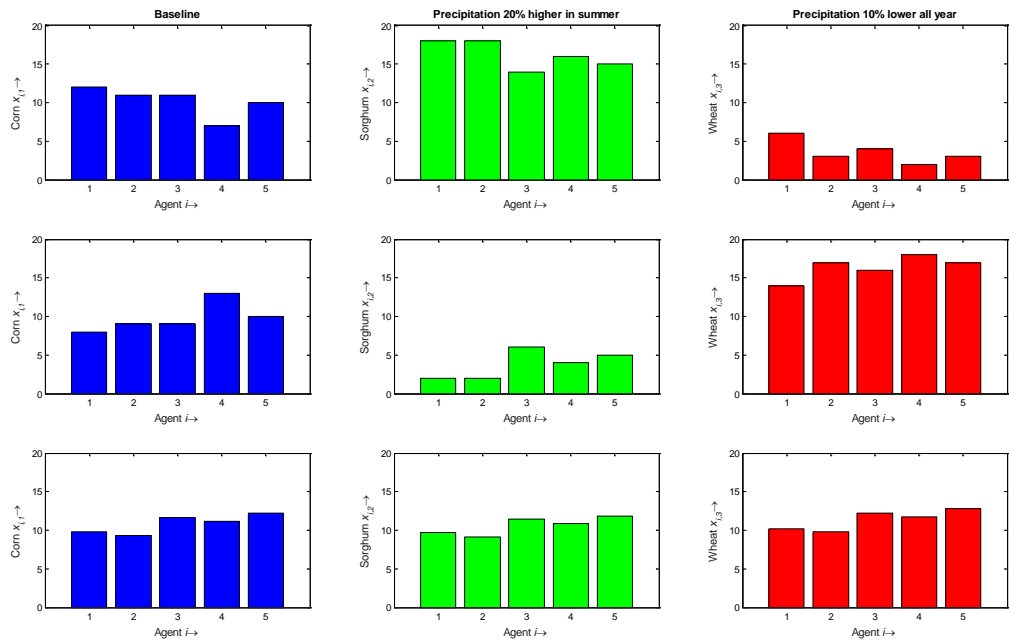

**Fig. 6.5.** Number of years a grain is grown under each scenario.

## 6.2 Socio-Economic Scenarios

The next study of the proposed MAS is to explore how changes in social and economic parameters in the future will affect groundwater use. The parameters that are affected in each case are shown in Table 6.1 below.

Table 6.1. Parameter changes under different socio-economic scenarios.

| Scenario | Parameters | | | | |
|---|---|---|---|---|---|
| | $G_0(t)$ | $p_k(t)$ | $g(t)$ | $c_k(t)$ | $IR_k(t)$ |
| Baseline | decrease | constant | constant | constant | constant |
| Population increase | decrease | increase | constant | constant | constant |
| Water use improvement | decrease | constant | constant | constant | decrease |

Fig. 6.6 shows the strategies of the five agents over time under this scenario. In anticipation of higher future crop prices, the agents try to limit water use during the initial periods by growing sorghum instead of corn in summer, switching to corn during the later stages. The wheat irrigation strategy also follows a similar trend with all five agents not growing any wheat during the first seven years, but subjecting the entire available land area to grow it during the last four years of the 20 year period. This observation is reflected in the utilities of the agents as shown in Fig. 6.7, where there is a steady increase in each agent's utility with time. Likewise in Fig. 6.8, the agents limited water use results in smaller declines in their groundwater heads during the initial years, with agents 4 and 5 showing marginal increases during the years 1 and 2.

The other scenario considers more efficient water use due to technological changes. The irrigation water requirement $IR_k$ for each crop $k$, which is obtained from Eqn. (4.3) is subjected to 2% decrease each year. Fig. 6.9 shows the strategies adopted by the agents each year. In comparison to Fig. 5.1, there is no marked change in the crops chosen to grow each year. Agents 1, 3 and 4 select to grow corn for only one more year in comparison to the baseline strategy; while agent 2 grows corn for 13 out of the 20 years in comparison to 11 years earlier. On the other hand, agent 5 does not show any change in its overall strategy by growing more corn. Thus, this model predicts that more efficient use of water for crop irrigation does not dramatically alter the irrigation patterns of the agents. The corresponding evolution in groundwater heads with time is shown in Fig. 6.10. The reduced groundwater requirement results in a somewhat slower decline in groundwater levels.

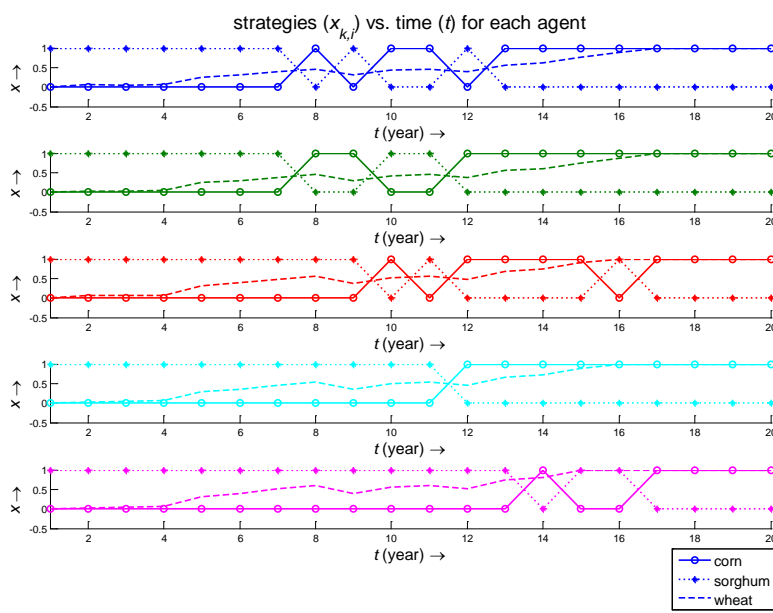

**Fig. 6.6** Strategies vs. year for each agent with population growth.

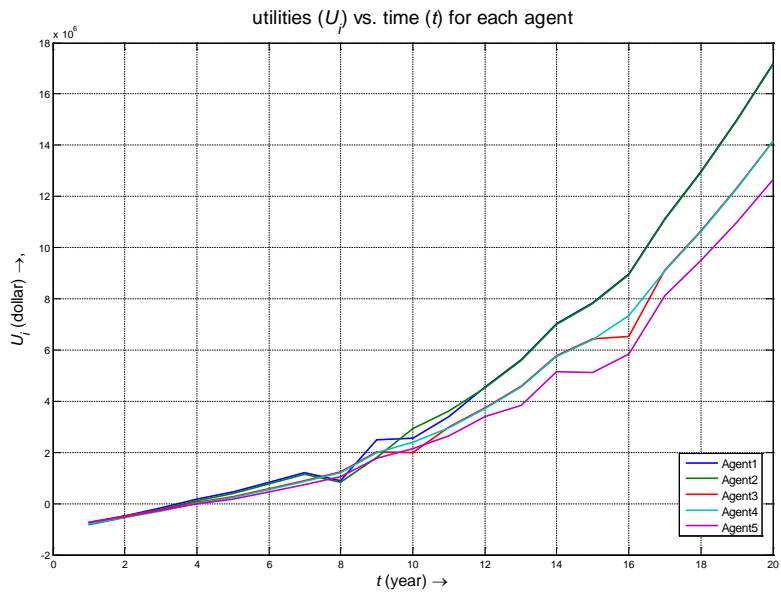

**Fig. 6.7** Utilities vs. year for each agent with population growth.

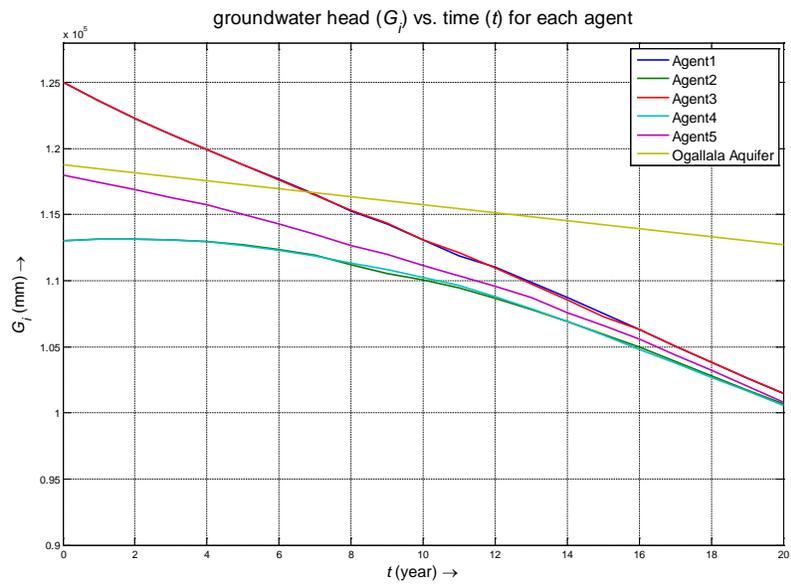

**Fig. 6.8** Groundwater head vs. year for each agent with population growth.

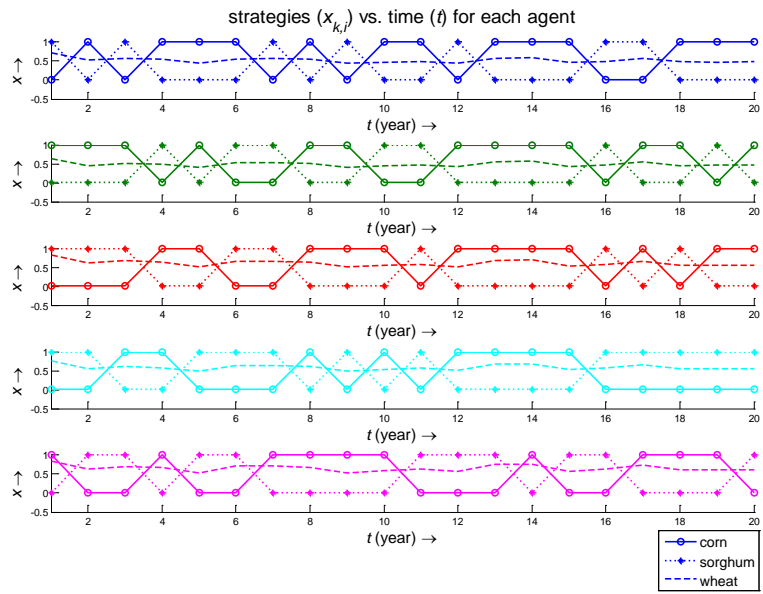

**Fig. 6.9** Strategies vs. year for each agent with more efficient groundwater use (2% per year).

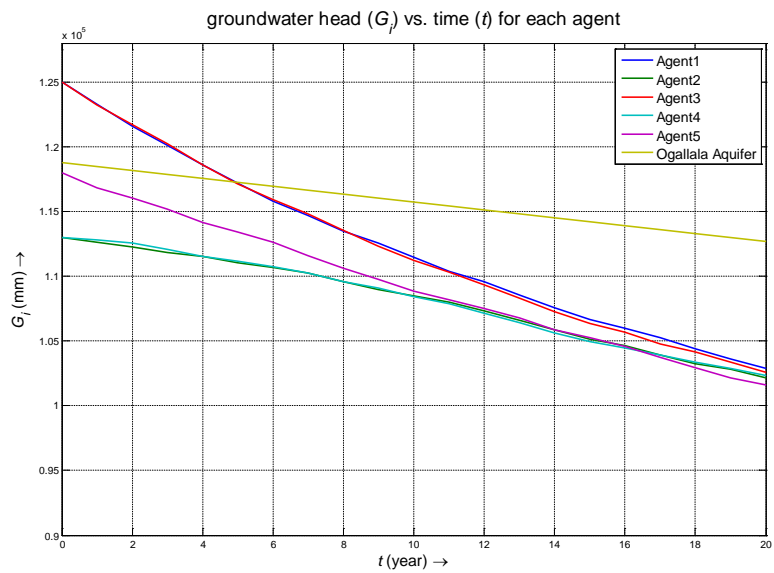

**Fig. 6.10** Groundwater head vs. year for each agent with more efficient groundwater use (2% per year).

# 7. Cooperative Groundwater Use

## 7.1 LEMA

The state of Kansas in 2012 has adopted a policy to enhance groundwater sustainability by authorizing the formation of Local Enhanced Management Areas (LEMAs). A LEMA is a local group of users who enter legally enforceable limits on extraction for water rights within a specified geographic boundary. The LEMA establishes its own limit on the total groundwater usage for a period of five years. Each irrigator is allowed to decide how to adjust its own irrigation strategy, such that the total groundwater extracted by it during this five-year period stays within the allowable limits imposed by the LEMA.

Since all simulations in this research have been performed for twenty years, the entire simulation can be divided into four non-overlapping 5-year periods. Let $\tau$ denote a set of five years representing any such 5-year period, say, $\tau = \{1,2,3,4,5\}$. The total volume of water pumped by agent $i$ at time $t \in \tau$ without any restrictions is given by,

$$W_i(t) = A_i \sum_k IR_k(t) x_{k,i}(t). \tag{7.1}$$

The agent's total water pumped during this period $\tau$ is,

$$W_i = \sum_{t \in \tau} W_i(t). \tag{7.2}$$

The LEMA limits the water consumption to a value $L_i$ which is a fraction of $W_i$. For instance $L_i = 0.9 W_i$, implying that the agent is required to lower its extraction by 10%.

## 7.2 Results

MAS simulations are carried out with $L_i = 0.95 W_i, 0.90 W_i, 0.85 W_i, 0.80 W_i, 0.75 W_i$ and $0.70 W_i$. There are four periods $\tau$ in these simulations. All five agents are assumed to form the LEMA.

The aggregate utilities for different values of $L_i$ are shown in Fig. 7.1. where the aggregate utilities are the totals over the 20-year duration. It is observed in Fig. 7.1 (top) that restricting groundwater depletion up to $L_i = 0.80 W_i$ results in steady increases in the utilities of agents 3, 4, and 5 which have smaller areas under irrigation, while those of agents 1 and 2 which have larger areas under irrigation is fairly consistent without any perceivable trend. When $L_i$ is reduced further, this trend is no longer observed.

Remarkably, the combined utility of all five agents, shown in Fig. 7.1 (bottom) also reveals an upward trend until $L_i = 0.80 W_i$. This trend can be explained as follows. When the agents with larger areas placed under irrigation limit their water use, due to groundwater flow across farms, the water heads at the remaining agents are higher. As a result, their costs of extracting groundwater (Eqn. (3.9)) is reduced.

Figure 7.2 shows the changed irrigation strategies at $L_i = 0.80 W_i$. As a result of the restricted groundwater use, the agents irrigate more sorghum instead of corn. Furthermore it is also seen that the strategies of agents 4 and 5 which have smaller areas under irrigation are shifted more than the others.

These results are encouraging, as they clearly show the effectiveness of LEMAs as a means to address the tragedy of the commons.

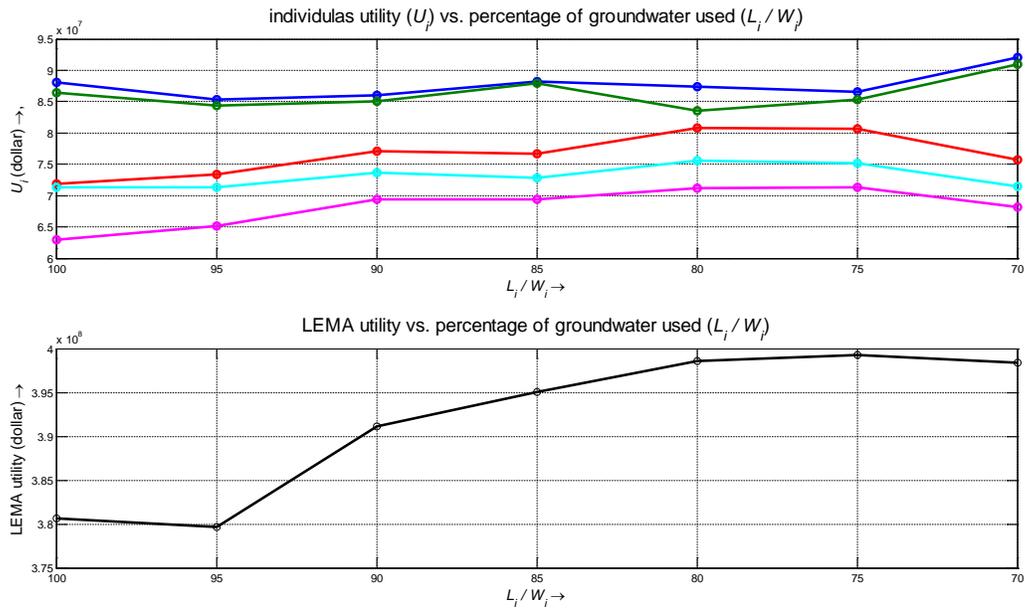

**Fig. 7.1** Utilities under different LEMA limits.

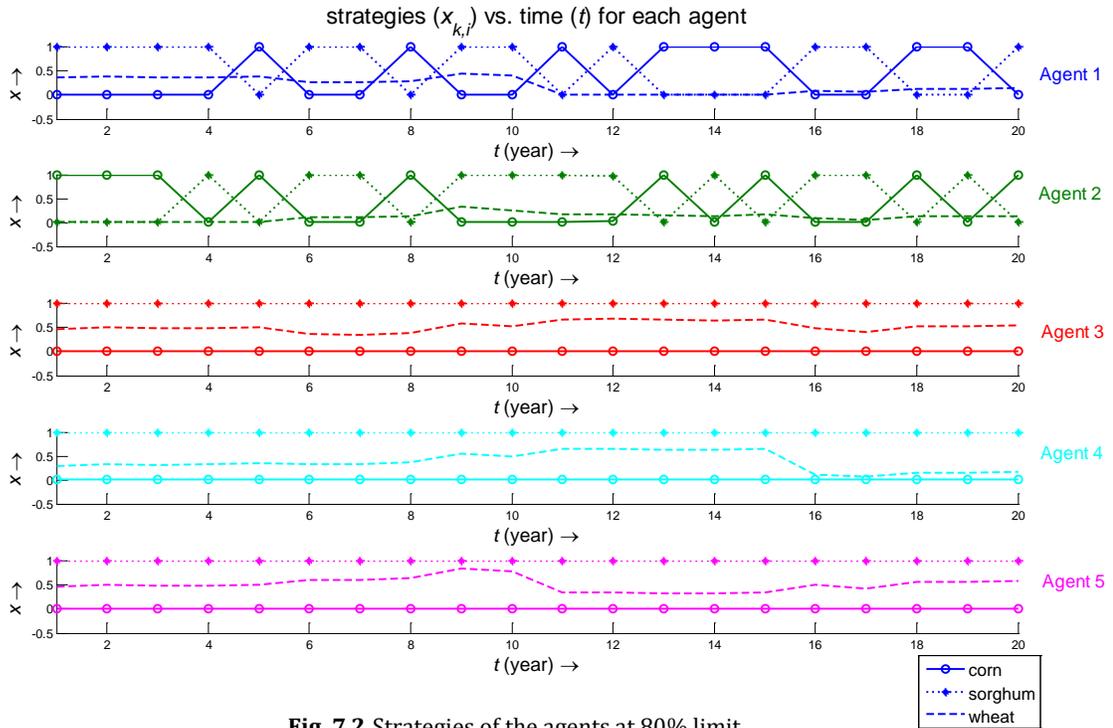

**Fig. 7.2** Strategies of the agents at 80% limit.

# 8. Conclusion

The MAS model in this research is shown to successfully simulate complex behavioral patterns shown in real irrigator agents. The following are the most significant aspects of real irrigation that is also seen in the MAS model .
- The agent strategies at Nash equilibrium incorporates crop rotation which is observed in real irrigation, where the agents alternate between the two summer crops, corn and sorghum.
- The MAS model exhibits the well-known phenomenon of economics of scale; agents with larger areas under irrigation are able to obtain slightly larger revenues per unit area under irrigation.
- The model is able to simulate various plausible climate-change scenarios.
- The model is able to simulate various plausible socio-economic scenarios.
- The model simulates cooperative groundwater use, showing that LEMAs will lead to better sustainable practices while increasing the aggregate payoffs of the irrigators.

Future research of this study can be conducted in the following directions.
- Extending the model to include more crop types.
- Enhancing the groundwater flow to include more geospatial features.
- Extending the model to larger sets of agents and incorporate coalition formation algorithms.
- Applying correlated equilibrium to obtain better irrigator payoffs under coordinated groundwater use.